\documentclass{article}[12pt]
\usepackage[pdftex]{graphicx}
\begin{document}
\title{ Remark on using quantum states prepared by the adiabatic quantum computation}
\author{Kazuto Oshima\thanks{E-mail: kooshima@gunma-ct.ac.jp}    \\ \\
\sl National Institute of Technology, Gunma College,Maebashi 371-8530, Japan}
\date{}
\maketitle
\begin{abstract}
We indicate that there are points to keep in mind in utilizing quantum states prepared by the adiabatic quantum computation.
Even if an instantaneous expectation value of a physical quantity for the adiabatically prepared quantum state is close to an expectation
value for the true vacuum, this does not assure us that the prepared vacuum is close to the true vacuum.
In general time average of the expectation value tend to systematically differ from the true value.   Using a simple model
we discuss how to diminish this systematic difference.
\end{abstract} 
\newpage
Recently, several  quantum systems are analyzed by quantum computers or quantum simulators with not so many qubits\cite{Martinez1}.
The quantum annealing\cite{Nishimori} gives a fundamental principle of the D-wave.
The adiabatic quantum computation, advocated by Farhi et.al.\cite{Farhi1} more than two decades ago, can also be carried out for quantum systems
with not so many number of qubits.   For some quantum field theories the adiabatic quantum computation have been used for
preparing ground states\cite{Honda, Okuda}.  After preparing ground states, it has been observed that an expectation value of certain
physical quantity varies significantly under a constant Hamiltonian\cite{Honda}. This oscillation
originates from the deviation of the prepared vacuum from the true vacuum.  It also has been observed that a time average of the
expectation value of the physical quantity systematically differs from the exact value computed by another method.

The purpose of this paper is to indicate that, even if  an instantaneous expectation value of a physical quantity for the approximate vacuum prepared by the adiabatic method is close to the expectation value for the true vacuum by chance, it is inevitable in general that an expectation value of a physical quantity for the approximate vacuum significantly oscillates in time around a point that slightly differs from an expectation value for the true vacuum. 
 We also discuss how to diminish this systematic difference.

According to the adiabatic quantum computation\cite{Farhi1}, we start from a simple Hamiltonian $H_{0}$ that has
a non-generate trivial vacuum.   We gradually change the Hamiltonian in time to a target Hamiltonian $H_{T}$ that we should analyze.
The quantum adiabatic theorem\cite{Born,Kato} assures us that the trivial vacuum of the initial Hamiltonian $H_{0}$ approaches the vacuum of the target
Hamiltonian $H_{T}$ if the change of the Hamiltonian is very moderate and there is a sufficient energy gap between
the vacuum and  excited states of the time varying Hamiltonian.  We simulate quantum adiabatic computation
using the quantum simulator by IBM for the simplest one-qubit case. We examine the quantum state prepared by the quantum adiabatic computation.

First, we choose the initial Hamiltonian ${\hat H}_{0}=-JZ, J>0$, and the target Hamiltonian ${\hat H}_{T}=-JX$.  The initial ground state
is $|0\rangle$ and the desired final state is $|+\rangle$.  The adiabatic Hamiltonian ${\hat H}_{A}(s)$ that connects ${\hat H}_{0}$ and ${\hat H}_{T}$
is given by
\begin{equation}
{\hat H}_{A}(s)=(1-s){\hat H}_{0}+s{\hat H}_{T}, \qquad  0 \le s \le 1,
\end{equation} 
where  for example $s={t \over T}, 0 \le t \le T$ for an adequate time period $T$. The quantum adiabatic computation starts at the time $t=0$ and finishes at the time $t=T$. 
For the ideal case, the ground state of the target Hamiltonian ${\hat H}_{T}$ has been prepared at the time $t=T$.  After the time $t=T$, we
observe a physical quantity under the constant Hamiltonian ${\hat H}_{T}$.  If the true ground state $|+\rangle$ has been prepared, 
an expectation value of the physical quantity is constant.  In the actual adiabatic quantum computation we have a quantum state that is slightly
different from the true vacuum $|+\rangle$.
Let us represent the state we have at the time $t=T$ as
\begin{equation}
|\psi(t=T)\rangle=\alpha|+\rangle+\beta|-\rangle, \quad |\alpha|^{2}+|\beta|^{2}=1,
\end{equation}
where $|\beta|^{2}$ is supposed to be small.
Under the total Hamiltonian  ${\hat H}_{T}=-JX$, this quantum state time develops as
\begin{equation}
|\psi(t)\rangle={\alpha}e^{iJ(t-T)}|+\rangle+{\beta}e^{-iJ(t-T)}|-\rangle,
\end{equation}
where we have set the Plank constant as $\hbar=1$ for simplicity.  For this state we measure the observable $Z$.
The expectation value of $Z$ time develops as
\begin{equation}
\langle \psi(t)|Z|\psi(t)\rangle=2|\alpha\beta|\cos(2J(t-T)+\theta),
\end{equation}
where the angle $\theta$ is defined by $\alpha\beta^{*}=|\alpha\beta^{*}|e^{i\theta}$.  Thus at the time $t=T$ we get the expectation value $2|\alpha\beta|\cos{\theta}$, which may be a good approximation of the desired value $\langle +|Z|+\rangle=0$ by chance.   The deviation, however, reaches up to $2|\alpha\beta|$ in the time development.  We can obtain the precise value by time averaging $\langle \psi(t)|Z|\psi(t)\rangle$ over a period.
Fig.1(a) shows one of our quantum simulation results. From the peak to peak value $4|\alpha\beta|$ we compute the varance $2|\alpha\beta|^{2}$ as $0.000730$, and
we get $2|\beta|^{2}=0.000730$.
By another simulation shots we observe $-X$ that commutes with the Hamiltonian. After the time $t=T$ the expectation value of $-X$ is almost constant(Fig.1(b)).   Its average is $-0.999320$ and its variance is $1.35 \times 10^{-9}$ for one of our simulation result with $10^{6}$ shots.   The value $-0.999320$ almost agrees with the previous value $-1+2|\beta|^{2}=-0.999270$.   

Second, we examine another simple one-qubit model.  We take the initial Hamiltonian as ${\hat H}_{0}=-JZ, J>0$, and we
take the target Hamiltonian as ${\hat H}_{T}=-JH$, where $H$ is the Hadamard gate.  We again observe the physical quantity $Z$. We represent
the eigenstates of the target Hamiltonian  ${\hat H}_{T}=-JH$ as $|h\pm\rangle$, where they satisfy $H|h\pm\rangle=\pm|h\pm\rangle$.
The explicit expressions of $|h\pm\rangle$ are
\begin{equation}
|h+\rangle={1 \over \sqrt{4-2\sqrt{2}}}(|0\rangle+(\sqrt{2}-1)|1\rangle),
\end{equation}
\begin{equation}
|h-\rangle={1 \over \sqrt{4+2\sqrt{2}}}(|0\rangle-(\sqrt{2}+1)|1\rangle),
\end{equation}
After the adiabatic state preparation process, the observable $Z$ time develops as 
\begin{equation}
e^{i{\hat H}_{T}t}Ze^{-i{\hat H}_{T}t}=e^{-iJ{H}t}Ze^{iJ{H}t}={1 \over \sqrt{2}}H-{1 \over \sqrt{2}}Y\sin{2Jt}+{1 \over 2}(Z-X)\cos{2Jt}. 
\end{equation}
At the time $t=T$ if we have a state $|\psi(t=T)\rangle=\alpha|h+\rangle+\beta|h-\rangle, |\alpha|^{2}+|\beta|^{2}=1$, instead of the 
desired state $|h+\rangle$, we have at a time $t(\ge T)$
\begin{equation}
\langle \psi(t)|Z|\psi(t)\rangle={1 \over \sqrt{2}}(1-2|\beta|^{2})+\sqrt{2}|\alpha\beta|\cos(2J(t-T)+\theta),
\end{equation}
where we have again set $\alpha\beta^{*}= |\alpha\beta^{*}|e^{i\theta}$. Since the physical quantity $Z$ does not anti-commute with 
the target Hamiltonian ${\hat H}_{T}=-JH$, the expectation value $\langle \psi(t)|Z|\psi(t)\rangle$ oscillate in time around the
value ${1 \over \sqrt{2}}(1-2|\beta|^{2})$ that slightly less than the desired value ${1 \over \sqrt{2}}=0.707107$.

Fig.2 shows one of our simulation results.  We have used the second order Suzuki-Trotter formula\cite{Trotter,Suzuki}.  In the result, a time average of $\langle \psi(t)|Z|\psi(t)\rangle$ is
0.706690, which we have computed from the average of the maximum value and the minimum value.   We can find the value $|\beta|^{2}$ from 
the variance of the values of $\langle \psi(t)|Z|\psi(t)\rangle$
in the range $t \ge T$.  In the result, the variance $|\alpha|^{2}|\beta|^{2}$ is $0.0003222$ and we find $|\beta|^{2}=0.0003223$.
Thus the expectation value of $Z$ is slightly improved to $0.707145$.  Thus the systematic error for the expectation value of
 $Z$ that is obtained from the time average has been diminished.

We have studied the quantum state preparation by the adiabatic quantum computation.  We have examined two simple 1-qubit models.
For the first case, the prepared quantum state is supposed to be a superposition of the true vacuum and the excited state. 
The expectation value oscillates in time around the expectation value of the true vacuum.  This is rather special case
that the physical  quantity anti-commute with the target Hamiltonian.   The second model will represent rather general case.
We observe the physical quantity that does not anti-commute with the target Hamiltonian.   In this case the time average of the expectation value
differs from the expectation value for the true vacuum.   We can diminish this difference from the time behavior of the expectation value.  
Although our models may be simple, our analysis would be useful to grasp properties of adiabatically prepared 
quantum states for more complicated systems, such as $(1+1)-$dimensional Schwinger model\cite{Honda}.

\newpage
Figure Captions\\
\\
Fig.1(a)\\
Simulation result of the adiabatic state preparation for the Hamiltonian
${\hat H}_{T}=-JX$ by IBM qasm-simulator.  We have started from
the ground state of ${\hat H}_{0}=-JZ$ and we have observed $Z$.   We have set $J=1$, the adiabatic time period $T=36$, and one time-step width
$\delta{t}={1 \over 8}$. The number of shots is $10^{6}$. The orange line represent the theoretical value.  After the time $T$, a time average over a period precisly leads to 0.
\\
\\
Fig.1(b) An expectation value of $-X$.   We have used another $10^{6}$ shots.
\\
\\
\\
Fig.2\\
Simulation result of the adiabatic state preparation for the Hamiltonian
${\hat H}_{T}=-JH$ by IBM qasm-simulator.  We have started from
the ground state of ${\hat H}_{0}=-JZ$ and we have observed $Z$.   We have set $J={\pi \over 4}$, the adiabatic time period $T=36$, and one time-step width
$\delta{t}={1 \over 24}$. The number of shots is $10^{6}$.  The orange line represent the theoretical value.  After the time $T$, a time average over a period slightly less than 
 the theoretical value ${1 \over \sqrt{2}}$.
\\
\newpage
{\quad }  \includegraphics[width=8cm]{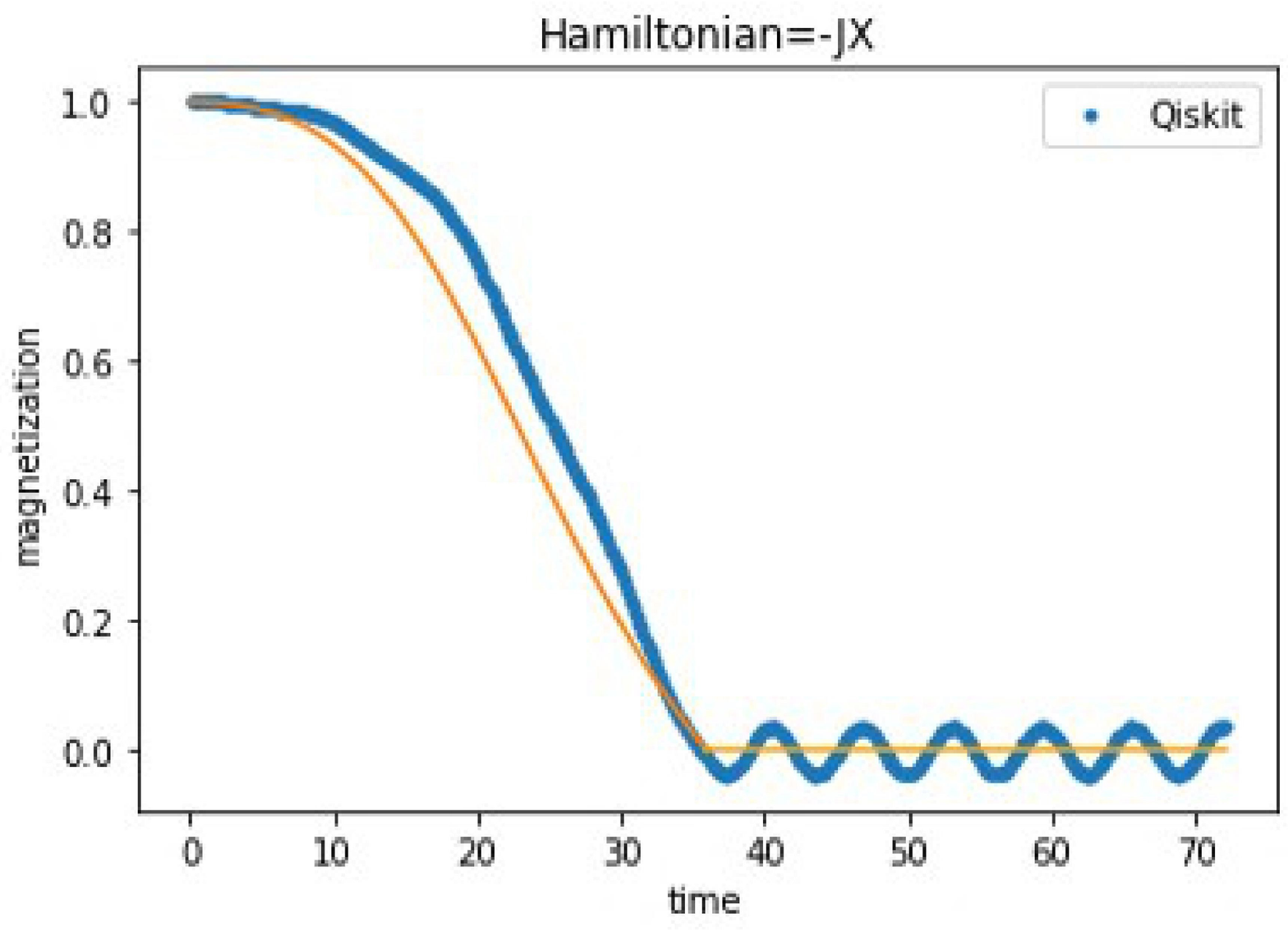}
Fig.1(a)\\
\vspace{10pt}
\\
{\qquad}  \includegraphics[width=8cm]{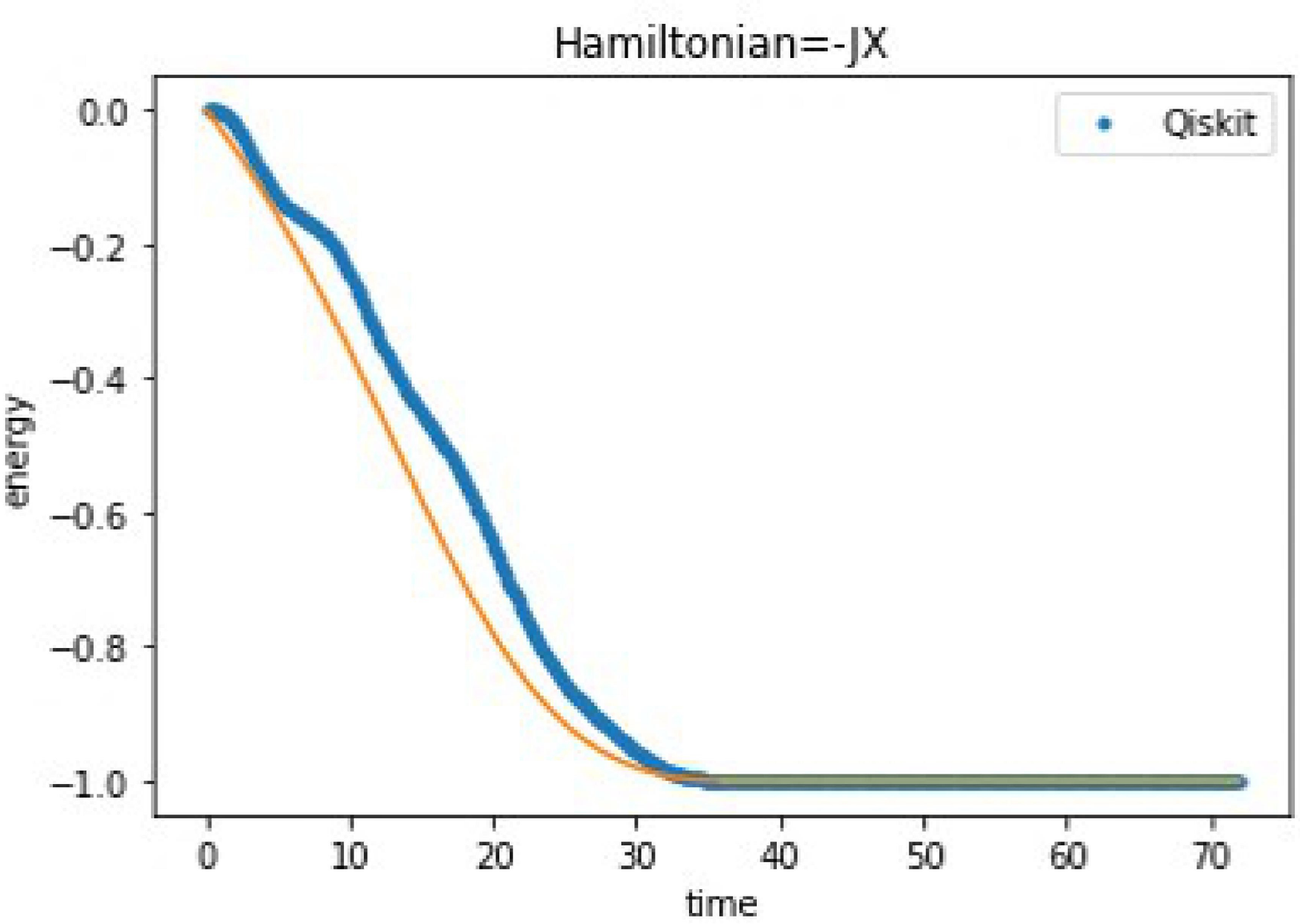}
Fig.1(b)\\
\vspace{10pt}
\\
{\qquad}  \includegraphics[width=8cm]{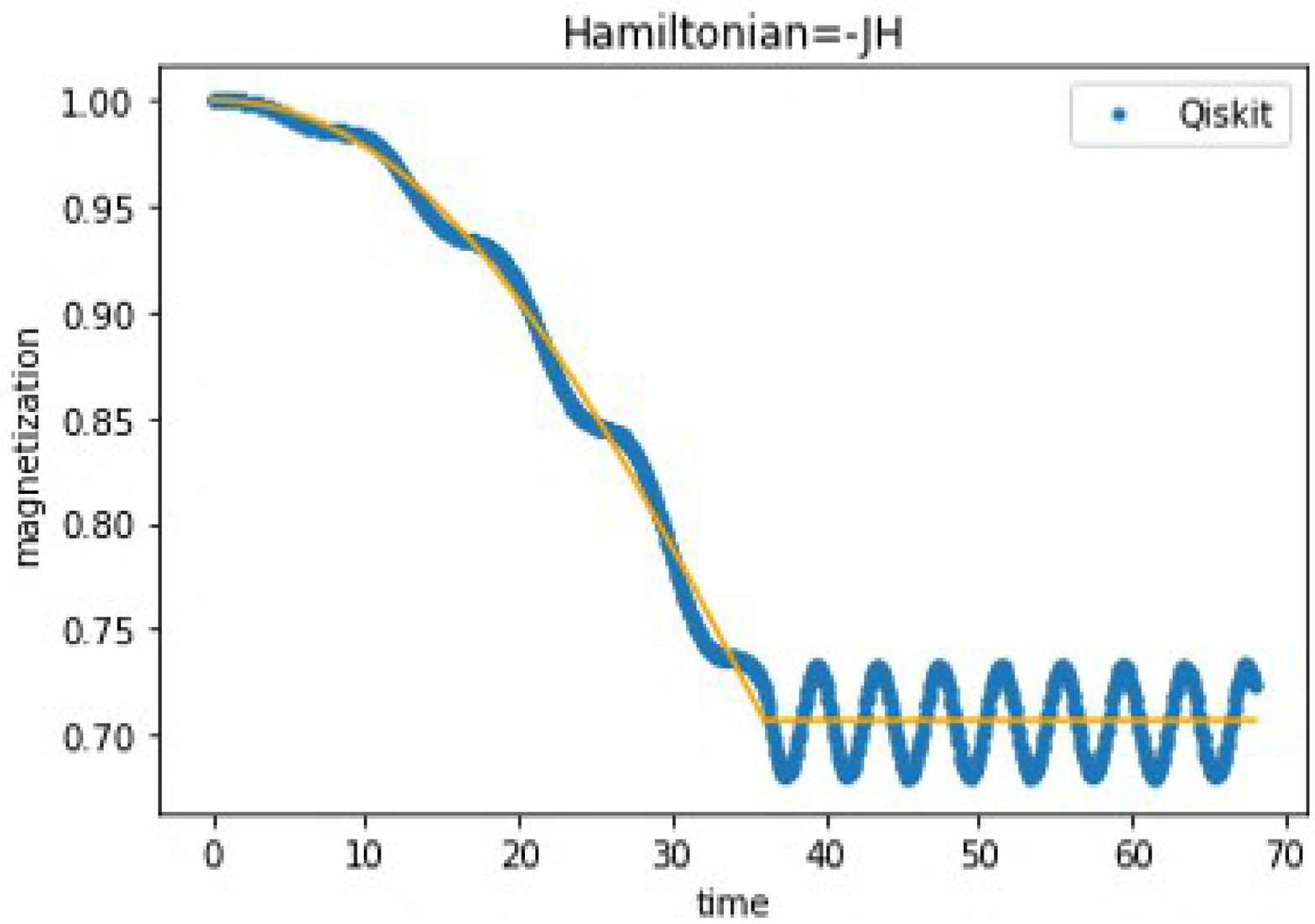}
Fig.2
\vspace{0pt}
\end{document}